# A SOLVER FOR MASSIVELY PARALLEL DIRECT NUMERICAL SIMULATION OF THREE-DIMENSIONAL MULTIPHASE FLOWS

Seungwon Shin[1,*], Jalel Chergui[2] and Damir Juric[2]

[1] Department of Mechanical and System Design Engineering

Hongik University

Seoul, 121-791 Korea

[2] Laboratoire d'Informatique pour la Mécanique et les Sciences de l'Ingénieur (LIMSI),

Centre National de la Recherche Scientifique (CNRS),

CNRS-UPR 3251, Bât. 508, Rue John von Neumann

Campus Universitaire d'Orsay

91405 Orsay, France

[*]Corresponding Author :

    Seungwon Shin, Ph.D

    Associate Professor

    Department of Mechanical and System Design Engineering

    Hongik University

    Sangsu-dong, 72-1, Mapo-gu

    Seoul, 121-791, Korea

    Phone: 82-2-320-3038

    FAX: 82-2-322-7003

    E-Mail: sshin@hongik.ac.kr

# ABSTRACT

We present a new solver for massively parallel simulations of fully three-dimensional multiphase flows. The solver runs on a variety of computer architectures from laptops to supercomputers and on 65536 threads or more (limited only by the availability to us of more threads). The code is wholly written by the authors in Fortran 2003 and uses a domain decomposition strategy for parallelization with MPI. The fluid interface solver is based on a parallel implementation of the LCRM hybrid Front Tracking/Level Set method designed to handle highly deforming interfaces with complex topology changes. We discuss the implementation of this interface method and its particular suitability to distributed processing where all operations are carried out locally on distributed subdomains. We have developed parallel GMRES and Multigrid iterative solvers suited to the linear systems arising from the implicit solution of the fluid velocities and pressure in the presence of strong density and viscosity discontinuities across fluid phases. Particular attention is drawn to the details and performance of the parallel Multigrid solver. The code includes modules for flow interaction with immersed solid objects, contact line dynamics, species and thermal transport with phase change. Here, however, we focus on the simulation of the canonical problem of drop splash onto a liquid film and report on the parallel performance of the code on varying numbers of threads. The 3D simulations were run on mesh resolutions up to $1024^3$ with results at the higher resolutions showing the fine details and features of droplet ejection, crown formation and rim instability observed under similar experimental conditions.

Keywords: direct numerical simulation, multiphase flow, parallel or distributed processing, interface dynamics, front tracking

# 1. INTRODUCTION

The area of numerical methods for free-surface and multiphase flow has seen major advances in the last two decades along with the computational resources to simulate them. The availability of massively parallel computing architectures promises access to simulations of demanding multiphase flows and spatio-temporal scales which are not possible on serial architectures. The central new feature to computational fluid dynamics when one is dealing with more than one phase is, of course, the phase interface. The presence of an interface and the ensuing non-linearities introduced to the problem pose great difficulties owing in part to the necessity to accurately compute surface tension forces and the need to provide a high fidelity treatment of these continuously evolving, deformable structures which often undergo drastic topology changes. Although many methods using various discretization approaches have been developed for handling phase interfaces (see [1] for an overview), two types of method have been most popular and successfully applied to various engineering applications. In one category are the Front Capturing type methods such as Volume-of-Fluid (VOF) [2], Level Set [3], and Phase Field [4] methods which represent the interface implicitly on a fixed Eulerian mesh. The second category includes Front Tracking type methods [5] which explicitly track the interface by using a separate Lagrangian discretization of the interface. Front capturing methods are generally simpler since no additional grid is required and interface advection is achieved by using high order schemes borrowed from methods developed for high speed compressible flow. Front tracking methods have shown generally the best performance in maintaining sharpness of the interface since the trajectories of the Lagrangian marker points which define the interface simply need to be integrated in time. On the other hand tracking methods introduce difficulties associated with a separate grid structure which is relied upon to faithfully model the complex dynamics and topology changes of physical interfaces.

Most implementations of parallel two-phase flow codes have focused on Front Capturing type methods owing to the relative ease of parallelizing Eulerian field data. One of the earliest parallel implementations involving phase interfaces was developed by George and Warren [6] for alloy solidification with the Phase Field method using the portable library DPARLIB and the Message Passing Interface (MPI) protocol for interprocessor communications. Wang *et al.* [7] used the processor virtualization technique to parallelize the Level Set method for the dendritic solidification problem. Processor virtualization maintains ideal load balancing on each processor by subdividing the computational domain dynamically thus improving cache management and communication. Sussman [8] developed a parallel adaptive two-phase flow solver based on the Level Set/Volume-of-Fluid (CLSVOF) method with adaptive mesh refinement (AMR) to further speed up the computations. Complex solid geometries were embedded on Cartesian grids for simulations of ship hydrodynamics. Fortmeier and Bucker [9] and Zuzio and Estivalezes [10] also developed parallel Level Set/AMR methods. Aggarwal *et al*. [11] used a domain decomposition technique to parallelize the Level Set method on simpler uniform grids for both single and two-phase flows including film boiling. The domain was decomposed unidirectionally with overlapping boundary data exchanged via MPI. Their parallelization performance was evaluated on a relatively small (2~16) number of processors. Recently, Banari *et al*. [12] implemented the Lattice Boltzmann method on massively parallel General Purpose Graphical Processing Units (GPGPU). The Smoothed Particle Hydrodynamics (SPH) method was also parallelized to investigate sloshing of a free surface. Popinet has developed a successful parallel AMR VOF flow solver called Gerris [13, 14] which has been widely used in many highly refined simulations of two-phase flows including an impressive numerical/experimental discovery of a von Kármán vortex street in the ejecta sheet of an impacting drop [15].

Front tracking has many advantages among them its excellent mass conservation, lack of numerical diffusion, and accuracy with which interfacial physics can be described on a subgrid level. It is found that tracking often does not require as highly refined grids. Tracking affords a precise description of the location and geometry of the interface and thereby the surface tension force (and other interface sources) can be very accurately computed directly on the interface. However, the interface is discretized by a mesh of moving triangular elements which adds complexity to the parallelization effort.

For the original Front Tracking method the cause of the difficulties lies in the need to logically connect the interface elements and bookkeep changes in connectivity during element addition, deletion or reconnection. In three-dimensions, the algorithmic complexity of a connected mesh of interface triangles increases dramatically, particularly for interface reconnection during topology change. Complicated procedures for bookkeeping interface connectivity and particularly for topological changes make the interface solver very difficult to parallelize. Bunner and Tryggvason [16] did however develop the first 3D parallel Front Tracking method to simulate a swarm of 91 separate bubbles on 8 nodes of an IBM SP2.

The technique our group has developed is the Level Contour Reconstruction Method (LCRM) [17-23], a hybrid Front-Tracking/Level Set method which retains the advantages of the classic Front Tracking method while doing away with the complexities of bookkeeping connectivity. Operations performed on each triangular interface element are independent of other elements, a feature which greatly facilitates parallel implementation.

Here we present our newly developed parallel solver for the simulation of two-phase incompressible flows based on the LCRM and its implementation on massively parallel computer architectures. The code combines: (1) high fidelity Lagrangian tracking of arbitrarily deformable phase interfaces, an accurate treatment of surface tension forces, interface advection and mass conservation with (2) the advantages of a Level Set distance function approach for periodic interface reconstruction which automatically and naturally handles breakup and coalescence. The solver is to our knowledge the first implementation of such a hybrid Front Tracking method on massively parallel architectures, running on up to 65536 threads and limited only be available machine resources. The code is general in that it can run on a variety of computer architectures from laptops to supercomputers. In this paper, details of the code structure and description of the parallel implementation of Lagrangian tracking and its coupling to the flow solver will be described. Even though we have conducted numerous quantitative benchmark tests of the code for a variety of single and two-phase flows, we turn our attention here to simulations of the splash of a drop onto a film of liquid, a flow with very large spatio-temporal extent where high performance is indispensable. A detailed quantitative physical study of the drop splash problem is not the purpose of this work but rather we focus on the parallel implementation and performance issues.

# 2. NUMERICAL FORMULATION

Due to the complex dynamics of interface motion, it is extremely difficult to formulate an accurate and general numerical scheme for multiphase flow simulations especially in three-dimensions. Among the variety of numerical techniques and grid structures for handling flows in the presence of interfaces, those which use a stationary underlying Eulerian/Cartesian grid for the primary velocity-pressure variables with an additional technique for interface treatment based either on the same Eulerian grid (thus capturing the interface) or based on a supplemental overlying moving grid (which tracks the interface) have become popular due to their relative simplicity and efficiency. Variants and hybrids of these basic Front Tracking or Front Capturing methods are widely used in multiphase applications. The basic idea in Front Tracking is that one tracks the motion of the interface explicitly using a separate discretized representation of the interface which moves with it. In Front Capturing one uses an additional field variable such as a volume function (VOF) or distance function (Level Set) to follow the interface motion implicitly. Nowadays, hybrid methods which retain only the desirable characteristics of both Front Tracking and Front Capturing approaches have become popular [24-27].

The Level Contour Reconstruction Method (LCRM) is one such hybrid which combines Front Tracking and Level Set techniques. The LCRM retains all of the usual Front Tracking features used to represent the interface with a triangular surface element mesh, calculate its surface tension and advect it. A major advantage of the LCRM over standard Front Tracking is that the interfacial elements are implicitly instead of logically connected. The LCRM periodically reconstructs the interface elements using a computed distance function field, a field such as the one used by the Level Set method, thus allowing an automatic treatment of interface element restructuring and topology changes without the need for logical connectivity between interface elements as was necessary in the original Front Tracking method. It is important to note that this distance function field plays no role whatsoever in the actual advection of the interface as it centrally does in the original Level Set method. We never need to solve an advection equation for the distance function. The LCRM thereby avoids the need for special corrective procedures to preserve mass such as Level Set reinitialization. An additional important benefit of the LCRM approach is that *all* operations are *local* to an individual triangular element independent of other elements. This principle of locality renders the LCRM particularly attractive for parallel computing since it carries over to distributed processing on local subdomains and thus its implementation on distributed processors is rather straightforward as will become clear in section 2.1 below.

Here we briefly describe the basic concept behind the LCRM and recent major improvements which include the use of high order interpolation, a vector valued distance function and tetra-marching in the interface reconstruction procedure. A more detailed description of this evolution can be found in [17-23]. The LCRM reconstructs the Lagrangian triangular interface elements by drawing constant contour surfaces of a distance function field as in Fig. 1 (shown for two-dimensions). Lines of constant contour can be drawn on the level contour field of the scalar distance function, $\phi$, at each reconstruction cell. These lines in each reconstruction cell share common end points and thus form implicitly connected surface elements across neighboring Eulerian grid cells. In the three-dimensional case, a rectangular grid cell (Fig. 2 (a)) will be divided into five tetrahedral reconstruction cells as in Fig. 2 (b). For each tetrahedral reconstruction cell, the interface reconstruction will be performed on cell faces similar to the 2D procedure above. After locating contour lines on the four faces of a tetrahedral reconstruction cell, the edges of contour faces can be obtained. Using the reconstructed edge lines, we can generate triangular elements as in Fig. 2(c). Since identical reconstructed edge lines are shared by neighboring reconstruction cells, all interface elements are implicitly connected without any logical connectivity. In order to ensure continuous reconstructed faces for the entire simulation domain, a tetra-marching orientation for the reconstructing tetrahedral cells is used as in Fig. 2(d).

## 2.1 Extended interface for parallel processing

In order to take advantage of the increased speed of today's modern high performance computing resources numerical methods for multiphase flows and the treatment of interfaces such as those discussed above must be adapted to multi-thread distributed processing and memory architecture.

Domain decomposition whereby the physical simulation domain (Fig. 3(a)) is subdivided into subdomains (Fig. 3(b)) each associated with a dedicated processing thread is relatively straightforward to implement for Eulerian/Cartesian meshes. Field variable data exchange takes place across neighboring subdomains via a boundary buffer zone. But with the existence of a moving Lagrangian interface grid as in the LCRM we need to apply an extended interface concept which uses additional buffer cells to store and exchange interface data necessary to the LCRM. As illustrated in Fig. 4, a subdomain is defined with two types of buffer zone: (1) a buffer zone for exchange of boundary data as before and (2) a new buffer zone for an extended interface which contains interface elements that extend beyond the physical subdomain boundary. There is an important difference in the way that some of the data is handled in these two buffer zones. In the boundary buffer zone, data is exchanged (copied) to adjacent subdomains in order to provide necessary boundary information overlap for the field variable solutions. In the extended interface buffer zone, each subdomain independently uses the Lagrangian interface information that is stored locally in its subdomain/buffer to perform interface operations and periodically apply the interface reconstruction procedure. The key advantage of the extended interface buffer is that interface operations are kept local to a subdomain and its buffer. Thus each subdomain/buffer handles tracking and operations on the Lagrangian interface elements (such as surface tension calculation) *independently* of the other subdomains. Since it is the boundary condition buffer that provides necessary field data such as the velocity field necessary for the Lagrangian element advection in the extended interface buffer and likewise for adjacent subdomain/buffers then the interface elements in the overlapping interface buffer zones will follow the same path independently. (This feature, which eases the task of parallelization greatly, can be viewed as having been inherited from the original LCRM philosophy of keeping operations local to an interface element or in this case local to a subdomain.) Finally since the interface must periodically be reconstructed, distance function values are also communicated in the extended interface buffer to adjacent subdomains in order to ensure interface connectivity across subdomains.

2.2 Solution procedure

Here, we will describe the basic solution procedure for the Navier-Stokes equations with a brief explanation of the interface method. Since our main focus in this work is on the numerical technique for distributed processing, details of the numerical solution method for the velocity, pressure and interface dynamics can be found in [17-23]. The governing equations for transport of an incompressible two-phase flow can be expressed by a single field formulation as follows:

$$\nabla \cdot \mathbf{u} = 0 \tag{1}$$

$$\rho\left(\frac{\partial \mathbf{u}}{\partial t} + \mathbf{u} \cdot \nabla \mathbf{u}\right) = -\nabla P + \rho \mathbf{g} + \nabla \cdot \mu(\nabla \mathbf{u} + \nabla \mathbf{u}^{\mathrm{T}}) + \mathbf{F} \tag{2}$$

where $\mathbf{u}$ is the velocity, $P$, the pressure, $\mathbf{g}$, the gravitational acceleration, and $\mathbf{F}$, the local surface tension force at the interface. $\mathbf{F}$ can be described by the hybrid formulation [18, 21]

$$\mathbf{F} = \sigma \kappa_H \nabla I \tag{3}$$

where $\sigma$ is the surface tension coefficient assumed to be constant and $I$ is the indicator function which is zero in one phase and one in the other phase. Numerically $I$ is resolved with a sharp but smooth transition across 3 to 4 grid cells. The Indicator function, $I$, is essentially a numerical Heaviside function and is generated using a vector distance function computed directly from the tracked interface [21]. Complete details of the procedure for obtaining the distance function directly from the interface geometry can be found in [18, 21]. $\kappa_H$ is twice the mean interface curvature field calculated on the Eulerian grid using:

$$\kappa_H = \frac{\mathbf{F}_L \cdot \mathbf{G}}{\sigma \mathbf{G} \cdot \mathbf{G}} \tag{4}$$

where

$$\mathbf{F}_L = \int_{\Gamma(t)} \sigma \kappa_f \mathbf{n}_f \delta_f (\mathbf{x} - \mathbf{x}_f) \mathrm{d}s \tag{5}$$

and

$$\mathbf{G} = \int_{\Gamma(t)} \mathbf{n}_f \delta_f (\mathbf{x} - \mathbf{x}_f) \mathrm{d}s \tag{6}$$

Here $\mathbf{x}_f$ is a parameterization of the interface, $\Gamma(t)$, and $\delta(\mathbf{x}\text{-}\mathbf{x}_f)$ is a Dirac distribution that is non-zero only when $\mathbf{x}=\mathbf{x}_f$. $\mathbf{n}_f$ is the unit normal vector to the interface and $ds$ is the length of the interface element. $\kappa_f$ is again twice the mean interface curvature but obtained from the Lagrangian interface structure. The geometric information, unit normal, $\mathbf{n}_f$, and length of the interface element, $ds$, in $\mathbf{G}$ are computed directly from the Lagrangian interface and then distributed onto an Eulerian grid using the discrete delta function. The details follow Peskin's [28] well known immersed boundary approach and a description of our procedure for calculating the force and constructing the function field $\mathbf{G}$ and indicator function $I$ can be found in [17-23].

The Lagrangian elements of the interface are advected by integrating

$$\frac{d\mathbf{x}_f}{dt} = \mathbf{V} \qquad (7)$$

with a second order Runge-Kutta method where the interface velocity, $\mathbf{V}$, is interpolated from the Eulerian velocity. Material properties such as density or viscosity are defined in the entire domain with the aid of the indicator function $I(\mathbf{x},t)$ as for example :

$$b(\mathbf{x},t) = b_1 + (b_2 - b_1)I(\mathbf{x},t) \qquad (8)$$

where the subscripts 1 and 2 stand for the respective phases.

The code structure consists of essentially two main modules: (1) a module for solution of the incompressible Navier-Stokes equations and (2) a module for the interface solution including tracking the phase front, initialization and reconstruction of the interface when necessary. The parallelization of the code is based on an algebraic domain decomposition technique. The code is written in the computing language Fortran 2003 and communications are managed by data exchange across adjacent subdomains via the Message Passing Interface (MPI) protocol. The Navier-Stokes solver computes the primary variables of velocity **u** and pressure P on a fixed and uniform Eulerian mesh by means of Chorin's projection method [29]. Depending on the physical problem, numerical stability requirements and user preferences, the user has a choice of explicit or implicit time integration to either first or second-order. For the spatial discretization we use the well-known staggered mesh, MAC method [30]. The pressure, and distance function are located at cell centers while the components of velocity are located at cell faces. All spatial derivatives are approximated by standard second-order centered differences. We use the multigrid iterative method for solving the elliptic pressure Poisson equation. As described in the next section, in the case of two-phase flow with large density ratio the now non-separable Poisson equation is solved for the pressure by a modified multigrid procedure implemented for distributed processors.

## 2.3 Parallel multigrid method for distributed computing

Another important issue in parallel computing for incompressible flow, in addition to the interface treatment, is solving the elliptic Poisson equation for the pressure. The projection method leads to a Poisson problem for the pressure which, for two-phase discontinuous density flows, is non-separable:

$$\nabla \cdot \left( \frac{1}{\rho} \nabla p \right) = S \qquad (9)$$

where, the density field $\rho$ is discontinuous since $\rho = \rho_L$ in the liquid and $\rho = \rho_G$ in the gas. The source term $S$ is a function of the non-projected velocities and interfacial tension.

For single-phase flow where $\rho_L = \rho_G$, the Poisson problem is separable and the conventional multigrid approach [31] especially for distributed computing has become quite attractive due to its near ideal efficiency compared with other iterative gradient based methods. However, in two-phase flows exhibiting large density differences between fluids the conventional multigrid technique becomes less efficient and often does not converge. Thus one of the most challenging issues, besides modeling the dynamics of the interface between the two fluid phases, is the efficient solution of the pressure equation for high density ratio $\rho_L/\rho_G$.

In our code we have developed a modified parallel 3D V-cycle multigrid solver based on the work of Kwak and Lee [32]. The solver incorporates a parallel multigrid procedure whose restriction and prolongation operators are not associated with each other, contrary to what is commonly used. This method has been successfully implemented to solve 3D elliptic equations where coefficients can be highly discontinuous [33]. The procedure can handle large density discontinuities up to density ratios of $O(10^5)$. The key features of the modified multigrid implementation can be summarized as follows:

① Cell centered second order finite difference approximation of equation (9).

② Harmonic approximation of the discontinuous coefficient $1/\rho$

③ Linear interpolation of the residual during the restriction process.

④ Cell flux conservation of the error on coarse grids during the prolongation process.

⑤ Parallel Red-Black SOR technique to relax the linear systems on fine grids.

⑥ Solution of the error using a parallel GMRES algorithm on the coarsest grid.

The first, second and fourth features are the most relevant parts of our multigrid solver. Without loss of generality, the relevant development can briefly be highlighted with a one-dimensional description. Denoting $\alpha = 1/\rho$, the cell center index $i = 1, ..., I_n = 2^{Ng-n}$, is defined on a set of Ng hierarchical Cartesian grids where $n = 0, ..., Ng$ and $h = h_n = L/2^{Ng-n}$. Here, $h$ represents the cell size of the grid and $L$ is the size of the domain. The pressure equation (9) is discretized on the finest grid ($n = 0$) by a second order cell centered finite difference technique as:

$$\alpha_{i-1/2}\left(p_{i-1} - p_i\right) + \alpha_{i+1/2}\left(p_{i+1} - p_i\right) = S_i h^2 \tag{10}$$

where, $\alpha_{i-1/2}$ and $\alpha_{i+1/2}$ are defined on the face centers of the cell i (Fig. 5). Since $\alpha$ is discontinuous, its value at $i+1/2$ and $i-1/2$ is better estimated by a harmonic approximation given by Eqs. (11) and (12).

$$\alpha_{i-1/2} = \frac{2\alpha_{i-1}\alpha_i}{\alpha_{i-1} + \alpha_i} \tag{11}$$

$$\alpha_{i+1/2} = \frac{2\alpha_{i+1}\alpha_i}{\alpha_{i+1} + \alpha_i} \tag{12}$$

Cell flux conservation gives an estimate of the error at the coarse grid cell faces, $e_{i-1/2}$ and $e_{i+1/2}$ as follows:

$$\frac{\alpha_{i+1/2}e_{i+1/2} - \alpha_i e_i}{\frac{h}{2}} = \frac{\alpha_{i+1}e_{i+1} - \alpha_{i+1/2}e_{i+1/2}}{\frac{h}{2}} \quad \text{East face} \tag{13}$$

$$\frac{\alpha_{i-1/2}e_{i-1/2} - \alpha_i e_i}{\frac{h}{2}} = \frac{\alpha_{i-1}e_{i-1} - \alpha_{i+1/2}e_{i+1/2}}{\frac{h}{2}} \quad \text{West face} \tag{14}$$

Of course, in three-dimensions, the analogous expressions must be evaluated at the 8 nodes, 12 edges and 6 faces of each cell of the coarse grid. The prolongation step then consists of using these values to interpolate the error at each cell center upward to the fine grids which produces a better initial guess of the error to relax. The efficiency of the multigrid method is measured by means of the so called average reduction rate $\gamma$ defined by Eq. (15).

$$\gamma = \left(\frac{R_{nc}}{R_o}\right)^{\frac{1}{nc}} \tag{15}$$

where, $nc$ is the number of cycles to convergence, $R_{nc}$ is the L2 norm of the residual at cycle $nc$ and $R_o$ is the initial L2 norm of the residual of each time step. $\gamma$ belongs to the interval [0, 1]. Values close to 0 indicate fast convergence whereas values close to 1 indicate slow convergence. The tolerance for convergence of the residual has been fixed to $10^{-9}$. Later in this article we investigate the performance of this modified multigrid procedure for a particular physical case.

# 3. RESULTS AND DISCUSSIONS

Two-phase flows inherently contain a broad spectrum of spatial and temporal scales whose concurrent numerical resolution requires today's massively parallel computing architectures. As part of the code development and testing process we have naturally conducted a variety of benchmark tests both with an interface (two-phase flows) and without an interface (single-phase flows) in order to evaluate the accuracy and the extent of applicability of our new parallel code. The results match those of previously existing serial procedures [17-23] and will not be reported here. The focus here will be on the new technology developed which allows for simulations not previously attainable on serial architectures. We present parallel performance results of massively parallel 3D simulations of the canonical drop splash problem where very fine grid resolution is necessary. The simulations involve the impact of a liquid drop onto a film or pool of liquid, a natural process occurring during rain showers, for example, and a process of interest in many engineering applications and fluid physics studies.

Several other groups have previously performed 3D numerical studies of droplet impact on a liquid film in order to analyze the evolution of the rich and varied hydrodynamic behavior of the impact, ejecta sheet, crown formation and droplet ejection [34-36]. In one often studied scenario involving drop impact on a shallow liquid film, the drop's downward momentum is rapidly transferred in a short distance into a radially outward and upward flow of liquid in the form of a thin vertical sheet which can undergo an unstable crown breakup and ejection of small droplets. A variety of intermediate stages are also observed. Another classic scenario is drop impact on a deeper liquid pool in which the rebound of the free surface after impact results in the upward ejection of a smaller satellite droplet from the center of impact. The full range of observed splash morphologies is enormous and depends critically on fluid properties, impact velocity and film thickness among other parameters. Adaptive mesh refinement (AMR) is often used in order to capture the details of the smallest interface structures as is done in the two-phase flow solver Gerris written by Popinet [13, 14]. Ideally the adaptive refinement technique concentrates computational effort only in regions where it is needed, for the refinement of the smallest details in a direct simulation, thus rendering accessible extremely highly resolved simulations for a given machine architecture. In practice, the algorithmic complexity makes it difficult to successfully implement AMR on massively parallel architectures although this problem will surely soon be overcome.

The approach in our present code uses a uniform mesh for its relative simplicity and ease of memory management which allows implementation on many thousands of threads of massively parallel machines such as the IBM BlueGene/Q. We focus on the qualitative physics of two drop splash scenarios and quantitative aspects of the parallel performance of the code.

**Drop Splash Simulations**

Here we report on the calculation intensive study of two regimes of drop splash simulation: the splash of a water drop onto a thin film and onto a deeper liquid pool of water. Even though the general behavior is commonly observed and widely studied experimentally especially since the pioneering work of Edgerton [37], the detailed physical process is quite difficult to capture from a numerical standpoint due to drastic changes of the interface dynamics and topology during the simulation. Nevertheless, a milestone in computation of a two-dimensional drop splash was achieved by Harlow and Shannon [38] in 1967.

Nearly five decades later we present in this work three-dimensional drop splash simulations performed on massively parallel supercomputers. In our simulations parallel performance was tested with weak scaling and showed reasonable efficiency on up to 32768 threads. The drop splash results showed physically appropriate behavior at all resolutions tested. However a qualitative match to experiments was obtained *only* with sufficiently high grid resolution, thus implying a need for sufficient computing capabilities (both algorithmic and hardware) to handle such high resolutions.

Details of the simulation geometry are shown in Fig. 6. We modeled two test cases which correspond roughly to case (b) (liquid pool) of Fig.1 in the experimental work of Okawa *et al*. [36], and case B (thin film) of Table 1 in the simulation study of Rieber and Frohn [35].

In our test case I, the deep liquid pool simulation, we set the Weber number($=\rho_L D V^2/\sigma$) to 580

and Ohnesorge number$(=\mu_L/(\sigma\rho_L D)^{1/2})$ to 0.0042. $\rho_L$ and $\mu_L$ represent density and viscosity of the liquid, respectively. *V* is the impact velocity, *D* is the diameter of the initial drop, and $\sigma$ is the surface tension coefficient. Initially, the center of the spherical droplet of diameter 1 mm is placed just above the horizontal surface of a liquid pool of depth H = 10 mm and with an impact velocity of 6.5 m/s. Water and air at atmospheric conditions are used as the working fluids; the liquid density and viscosity, are 1000 kg/m$^3$ and 1.137×10$^{-3}$ kg/m/s, respectively; the corresponding values for gas are 1.226 kg/m$^3$ and 1.78×10$^{-5}$ kg/m/s, respectively. The surface tension coefficient is assumed to be constant and equal to 0.0728 N/m.

For our test case II, the thin film splash, case B of Table 1 in Rieber and Frohn [35], the simulation conditions are similar except we use a film thickness of 0.7 mm, an initial droplet diameter of 7 mm, and impact velocity of 2.13 m/s. This gives a Weber number of 436 and an Ohnesorge number of 0.0016.

**3D Simulation Results**

**Case I:** Full simulations were performed on 4 different grid resolutions of 128$^3$, 256$^3$, 512$^3$, and 1024$^3$. Snapshots of the interface evolution at specific times with three different resolutions of 128$^3$, 256$^3$, and 512$^3$ are shown in Figs. 7 to 9. As can be seen from Fig. 7, the initial impact creates a crater at the center with a somewhat smooth rim which spreads out radially. The crater collapses inward forming a small central peak. With increased resolution to 256$^3$ (Fig. 8), the interface details become noticeably sharper. The outward spread of the rim after the initial impact is in a much narrower band and the crater collapse and rebounding central upward jet is quite noticeable. But still this jet is not quite high in comparison with the experimental results. The results with 512$^3$ resolution (Fig. 9) show a drastically different behavior compared to the other lower resolution cases. With sufficient resolution, the crater collapse and upward jet formation is correctly simulated. Furthermore the strong upward jet leads to breakup and generation of a satellite droplet, providing a good qualitative match to the behavior seen in Fig. 1(b) of Okawa *et al* [36]. These results indicate that sufficient resolution is imperative to achieve even a qualitatively correct physical solution.

In Fig. 10, the maximum surface location was plotted vs. time for each grid resolution. As can be seen in the figure, the behavior of the interface changes drastically at and above $512^3$ grid resolution. The grid resolution of $1024^3$ shows an almost similar behavior compared to the $512^3$ resolution. Due to the enormous number of computing hours consumed by the $1024^3$ case running on 32768 threads we were only able to run the $1024^3$ resolution until about 20 ms. However, the $1024^3$ and $512^3$ resolutions show similar results and we can conclude that grid convergence for this drop splash case has been essentially reached with a $512^3$ grid.

**Case II :** The fully three-dimensional nature of the interface dynamics can be seen in test case II in Figs. 11-13. In this simulation resolved on a $512^3$ grid, a relatively large droplet compared to test case I impacts a thin liquid film. A radially outward projecting ejecta structure can clearly be seen at a time of 1.3 ms. As the drop further merges with the thin film this structure is pushed outward and upward into a rim structure which increases in radius and height until the upper portion of the rim becomes unstable to a characteristic wavelength around 7 ms from which smaller droplets are then ejected. The crown and wavelength selection dynamics have been extensively studied and several physical mechanisms proposed [39-46]. The crown exhibits a form very close to experimental results. It is also remarkable to observe an extremely fine capillary wave structure extending downward from the rim, particularly visible on the inside surface of the rim below the crown after 14 ms. Velocity vectors on the mid-plane are plotted in Fig. 12 with a closeup at t=21.99 ms shown in Fig. 13. At early times the high speed ejecta sheet is clearly visible. Recent high resolution experiments and simulations have shown the existence of an intricate small scale von Kármán street in the ejecta sheet [15]. At later times the expanding vertical rim sheet creates a vortex of air flowing from outside the sheet over the top of the crown to the inside of the rim. For this test case, even higher resolution will be necessary to obtain quantitatively accurate interface dynamics but the result shows qualitatively reasonable results with the current resolution. This result builds confidence in the ability of our code to simulate complex inherently three-dimensional interface dynamics with severe topological changes on massively parallel machines.

**Parallel Performance**

The simulations reported in this work were run on BlueGene/Q supercomputers whose construction is analogous to nested dolls. The basic architectural framework is a rack which contains 32 node cards. Each node card contains 32 compute cards. Each compute card is composed of 16 cores (chips sharing 16 GB of DDR3 memory) and each core has 4 threads. If all 64 threads of a compute card are used (the hyperthreading option which we use in our computations) then the memory of each thread is limited to 256 MB.

We use the common parallelization strategy of domain decomposition in which the global simulation domain is divided into a desired number of spatial subdomains. Each subdomain is assigned a dedicated thread. The computations have been performed on up to 32768 threads of a BlueGene/Q supercomputer. Each thread is mapped to an MPI process. At the hardware level on the BlueGene/Q machine each subdomain/MPI process/thread has access to 256 MB of memory. In the simulations performed here the simulation domain is divided spatially into either 512, 4096 or 32768 subdomains whose boundary data may have to be communicated across adjacent subdomains and synchronized globally at appropriate times during the computation.

There are two generally prevalent ways to evaluate the performance of parallel codes. The first is called "heavy scaling": varying the number of threads and keeping the global mesh resolution fixed. The second is called "weak scaling": varying the number of threads and keeping the local (subdomain) mesh resolution fixed. Here, we are more concerned with the latter. This is mainly for two reasons: (1) on the Bluegene/Q machine each thread has access to a relatively small memory of 256 MB and (2) we are more interested in obtaining the finest global mesh resolution by increasing the number of threads (subdomains) instead of by increasing the local mesh resolution. We find that a subdomain grid of 32×32×32 seems optimal. Thus, for example, using 8192 = 16×16×32 threads gives a global resolution of 512×512×1024.

Timings of the simulation are measured where the speedup $S(N_t)$ and efficiency $E(N_t)$ are deduced in weak scaling according to the formulas given by Eq. (16).

$$S(N_t) = \frac{N_t T(N_r)}{N_r T(N_t)} \qquad E(N_t) = \frac{N_r}{N_t} S(N_t) \qquad (16)$$

Table 1 shows the overall performance of the code in weak scaling for three runs with number of threads $N_t$ = {512, 4096, 32768}. Also shown are the efficiencies which fall in the range [0, 1]. $N_r$ = 512 is an arbitrarily chosen reference number of threads. $T(N_t)$ is the measured wall clock time to reach 1 ms of physical time normalized by the number of time steps taken for each of the runs. In weak scaling, we fixed the local mesh size to $32^3$ regardless of the number of threads in order to keep the computational load per thread constant.

With no overhead for the cost of communication and synchronization between threads, the wall clock time $T$ should remain constant, which is never the case in real parallel computations. Data exchange and synchronization overheads incur a speedup and efficiency cost and will generally affect the overall scalability on massively parallel machines depending on the amount and the frequency of data exchanged between MPI processes. In fact, for a larger number $N_t$ of threads, the wall clock time $T$ slightly increases with a doubling of the global mesh resolution as can be seen in table 1. The speedup index, $S$, indicates how fast the simulation can run with respect to a reference number $N_r$ = 512 of threads. As can be seen in table 1, the code is 5.6 times faster and maintains an efficiency E of 70% up to 4096 threads. However, the efficiency becomes lower (E = 48%) even though the code is 31 times faster with 32768 threads. This is partly because the number of communication links between subdomains has been increased significantly by a factor of 384. The ratio of subdomains has increased by a factor of 64 from 512 to 32768 threads in the three-dimensional geometry since each subdomain has 6 neighbors.

We are also interested in the repartition of the computational time during one time step among the two main routines: the Navier-Stokes solver and the interface solver. The time spent in the Navier-Stokes module is determined mainly by the iterative parallel solvers GMRES and Multigrid which dynamically adapt the number of iterations to convergence in each time step during the simulation. The interface solver involves computation of the interface advection, the surface tension force and the Lagrangian mesh reconstruction. In a normal time step without reconstruction the Navier-Stokes solver takes 80% of the CPU time while the interface solver takes 20%. The situation is different during a time step that invokes the Lagrangian mesh reconstruction procedure where it can be seen that the reconstruction procedure occupies the most computational time. The time spent in the interface solver jumps to 70% versus 30% for the Navier-Stokes solver. However, the reconstruction, although expensive in one time step, is not performed in every time step but takes place only episodically under fixed criteria. A user defined automatic control routine for the frequency of reconstruction has been built in and the reconstruction is performed when the interface has traveled far enough (usually one grid size). This control routine greatly reduces the overall cost of the reconstruction by invoking it only when necessary. Alternatively, the user can define the frequency of reconstruction. In this case, the interface reconstruction process is activated periodically after a given number of time steps, usually on the order of every 100 time steps for most general cases. Future versions of our code will contain optimization improvements to the reconstruction routine which should greatly reduce its cost. Table 2 gives an average of the CPU time repartition between the two routines for one time step with and without reconstruction of the Lagrangian mesh interface.

The Navier-Stokes solution involves the solution of a non-separable Poisson equation for the pressure which, for large density ratios between fluid phases, can be quite ill-conditioned and is usually the costliest portion of the Navier-Stokes solver. We are interested in the performance of the multigrid method used for the pressure calculation as measured by the average reduction rate $\gamma$ as defined by Eq. (15). Figures 14 (a) and (b) show the time history of the average reduction rate during the simulation of both test case I and II respectively. The reduction rate is a normalized measure of multigrid efficiency. Values close to 0 indicate fast convergence whereas values close to 1 indicate slow convergence. As shown in figure 14 (a), for test case I the convergence rate of the multigrid drops suddenly ($\gamma = 0.5$) at t = 0.06 sec when the satellite droplet redescends and coalesces with the free surface. Also, as shown in Fig. 14 (b) for test case II, the convergence rate slows ($\gamma = 0.5$) in the time interval $t$ = [0.0075 s, 0.025 s] during droplet detachment from the crown and coalescence.

# 4. CONCLUSION

We have presented some of the capabilities of our new solver for massively parallel simulations of three-dimensional multiphase flow. The parallel implementation of the LCRM hybrid Front Tracking/Level Set interface method is shown to provide high quality solutions to the difficult problem of drop splash onto a liquid film where the air/water interface undergoes rapid and complex dynamics including multiple topology changes. This solver is to our knowledge the first implementation of a Front Tracking hybrid method on massively parallel computing architectures. A quantitative investigation of the details of the many physical mechanisms observed in the drop splash problem is beyond the scope of our study. Since our computing resources on the massively parallel BlueGene/Q machines available to us were limited, we focused our attention on simulations of the drop splash problem for two regimes: drop impact on a thin liquid film and on a deeper liquid pool. The 3D simulations run on mesh resolutions up to $1024^3$ display the fine details and features of droplet ejection, crown formation, rim instability and capillary waves on the rim observed under similar experimental conditions. The stark difference in qualitative results between simulations on different mesh resolutions is remarkable and points to the need to achieve a minimum sufficient resolution to capture even the overall qualitative features of the drop and splash dynamics. We report on the parallel performance of the code in weak scaling where the local subdomain resolution was maintained while the number of threads and the global resolution increased. Reasonable speedups and efficiencies for this difficult case were obtained and highlight the fact that parallelization is indispensable in achieving such high grid resolution simulations. Simulations on a $1024^3$ mesh with 32768 threads ran 31 times faster than the same case on a $256^3$ mesh with 512 threads. Good performance was also seen for the parallel multigrid method used to solve the pressure Poisson problem with strong density discontinuity. Since the pressure solution is usually a very costly part of the solution of incompressible flows we paid particular attention to the design of an efficient and robust parallel implementation of multigrid for strongly discontinuous density.

# ACKNOWLEDGEMENTS

This work was supported by: (1) the Partnership for Advanced Computing in Europe (PRACE), (2) the Institute for Development and Resources in Intensive Scientific Computing (IDRIS) of the French National Center for Scientific Research (CNRS), coordinated by GENCI (Grand Équipement National de Calcul Intensif) and (3) the Basic Science Research Program through the National Research Foundation of Korea (NRF) funded by the Ministry of Education (2012R1A1A2004478). Data post processing is performed with the ParaView visualization platform. We thank M. Firdaouss, L. Tuckerman, and P. Le Quéré for helpful discussions.

Table 1. Performance in weak scaling on the BlueGene/Q machine.

| Global mesh size | $N_t$ | $T(s)$ | $S$(/Ideal) | $E$ |
|---|---|---|---|---|
| $256^3$ | 512 | 14 | 1.0(/1) | 1.00 |
| $512^3$ | 4096 | 20 | 5.6(/8) | 0.70 |
| $1024^3$ | 32768 | 28 | 31.0(/64) | 0.48 |

Table 2. Percent CPU time repartition between the interface solver and the Navier-Stokes solver for one time step.

| Interface without reconstruction / Navier-Stokes | Interface with reconstruction / Navier-Stokes |
|---|---|
| 20% / 80% | 70% / 30% |

# FIGURE CAPTIONS

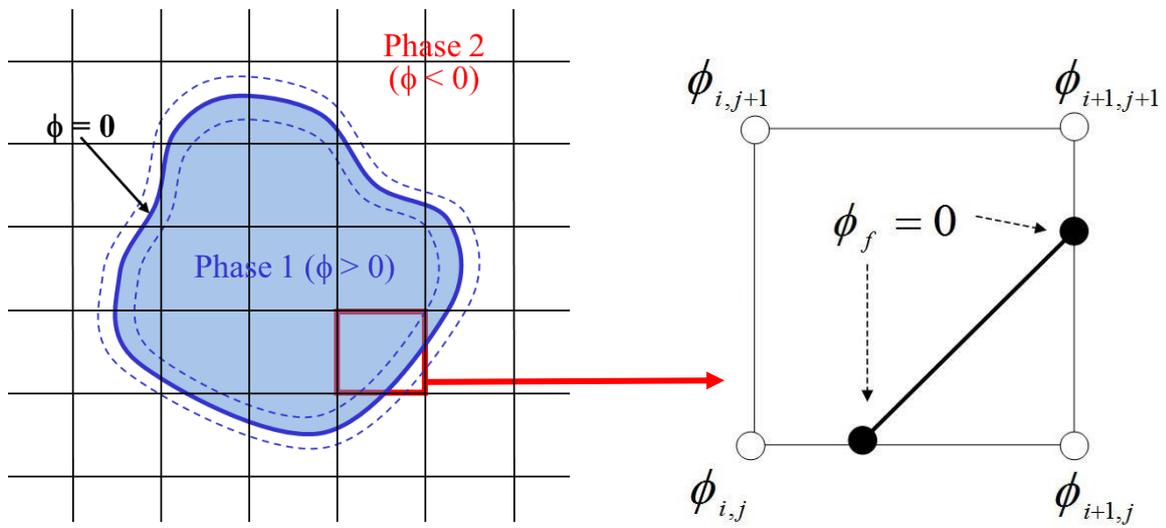

Fig. 1

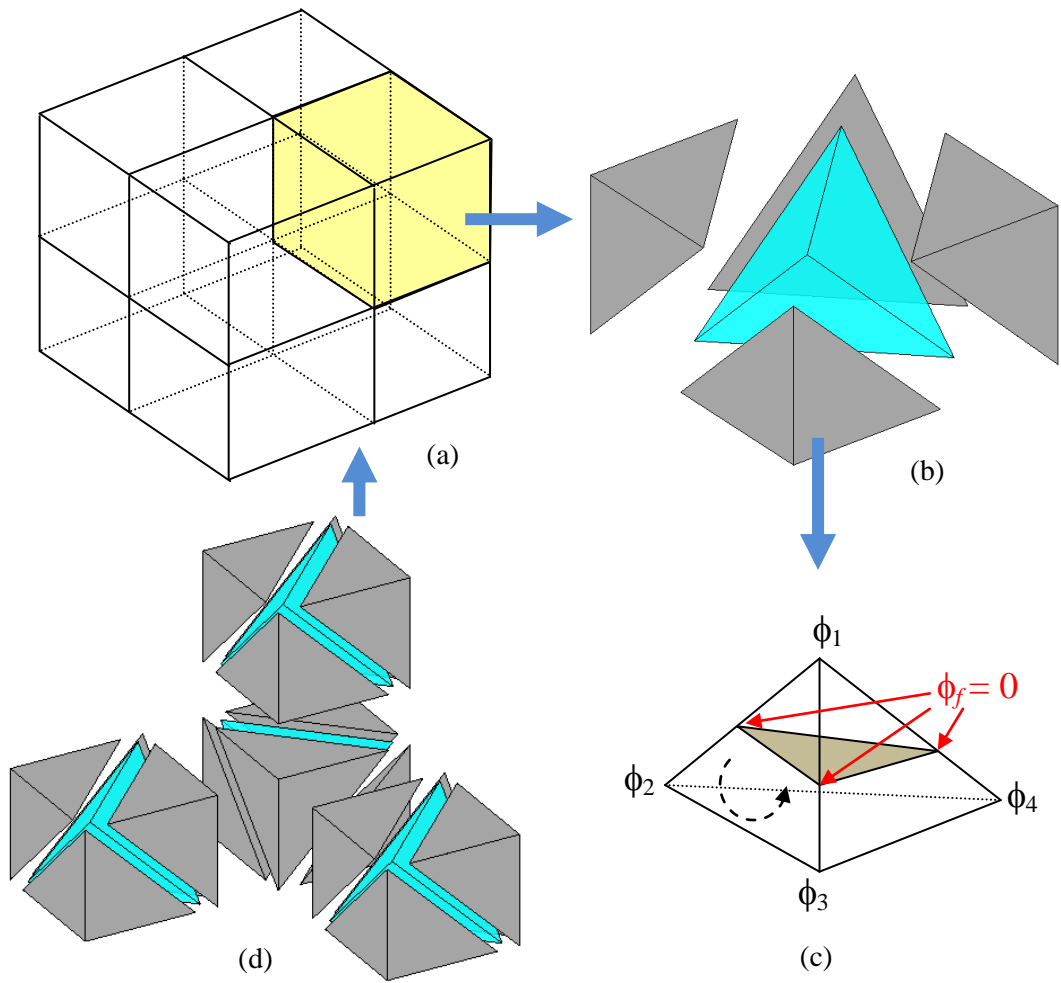

Fig. 2

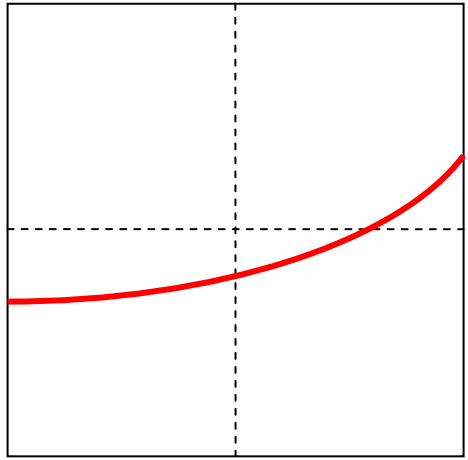

(a)

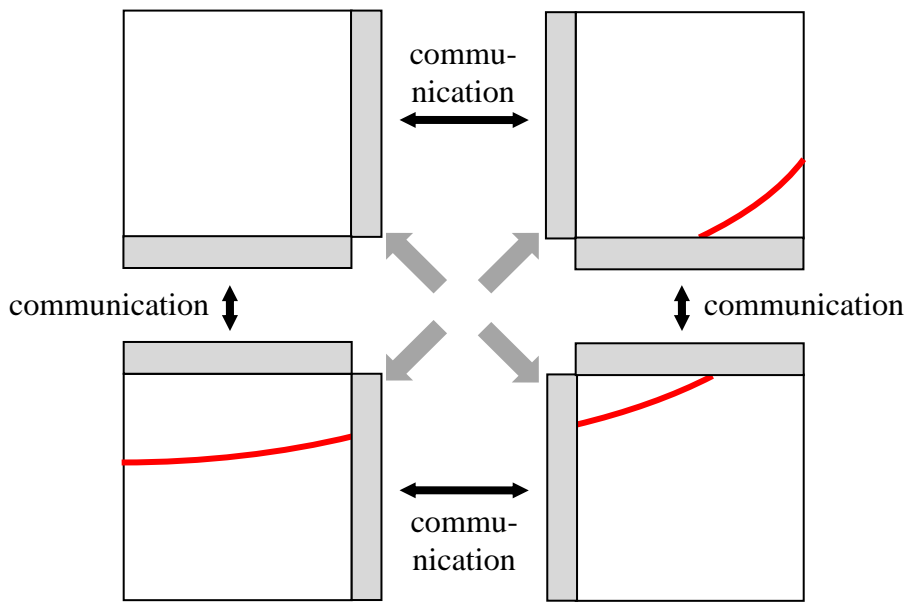

(b)

Fig. 3

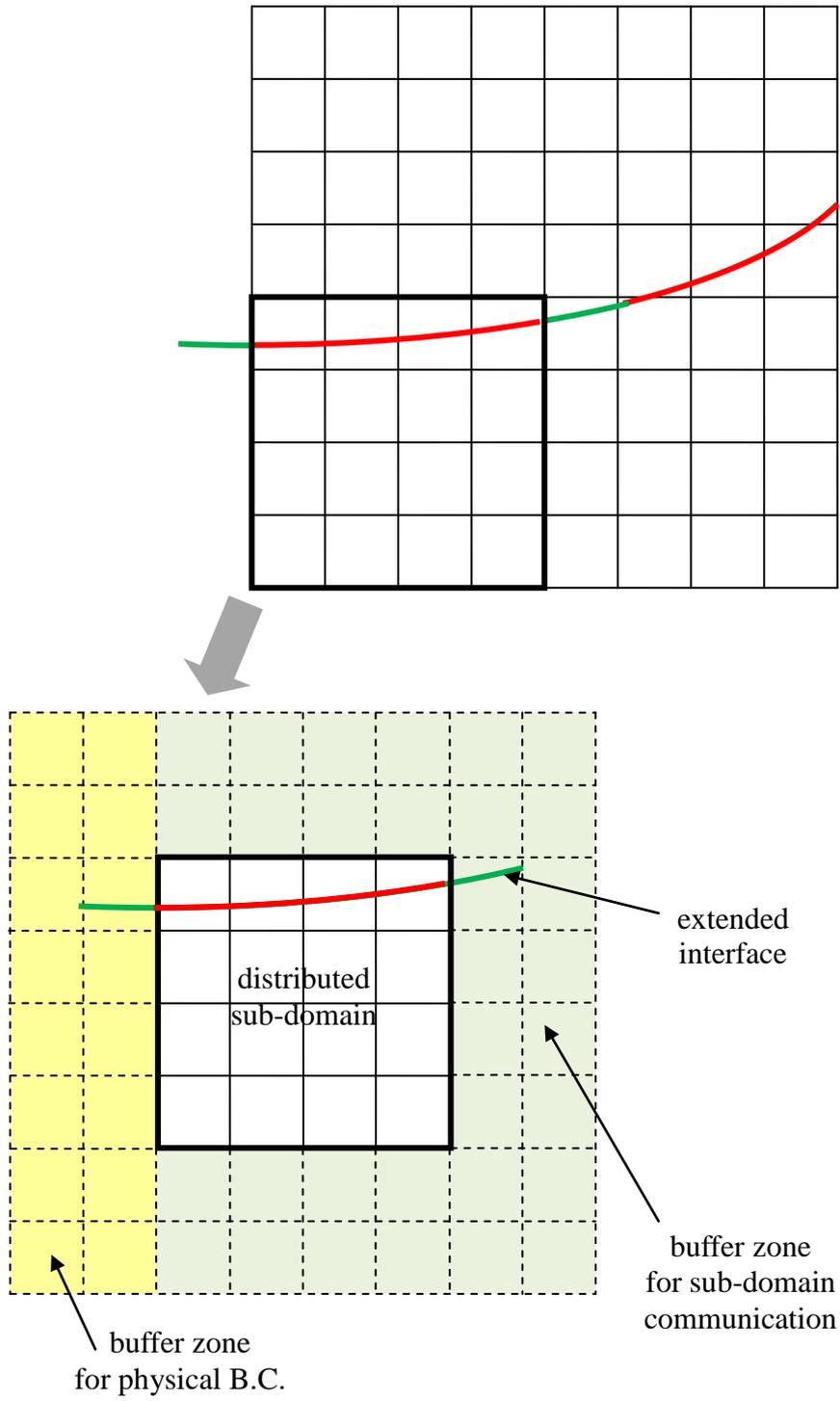

Fig. 4

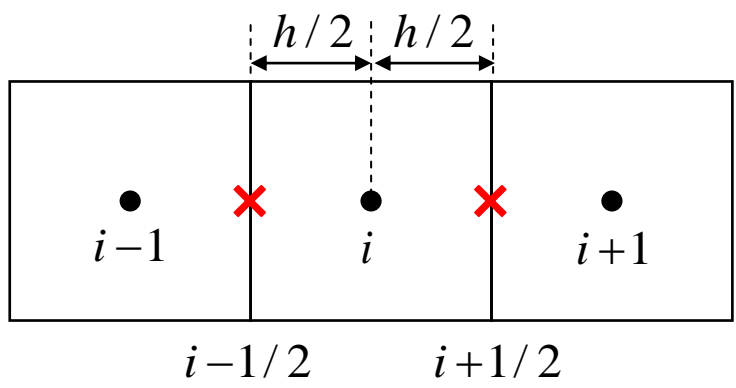

Fig. 5

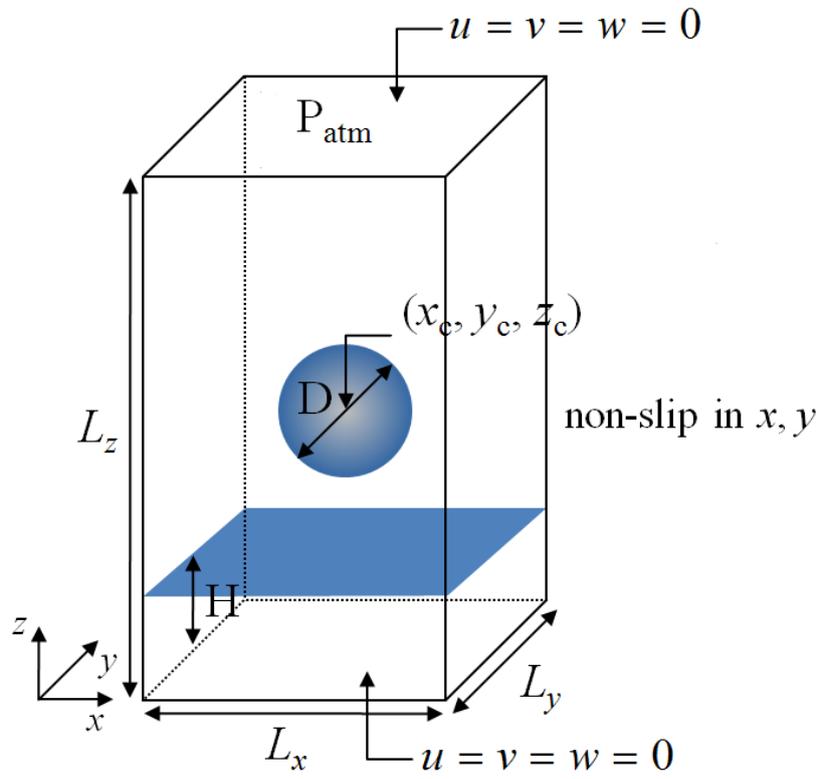

Fig. 6

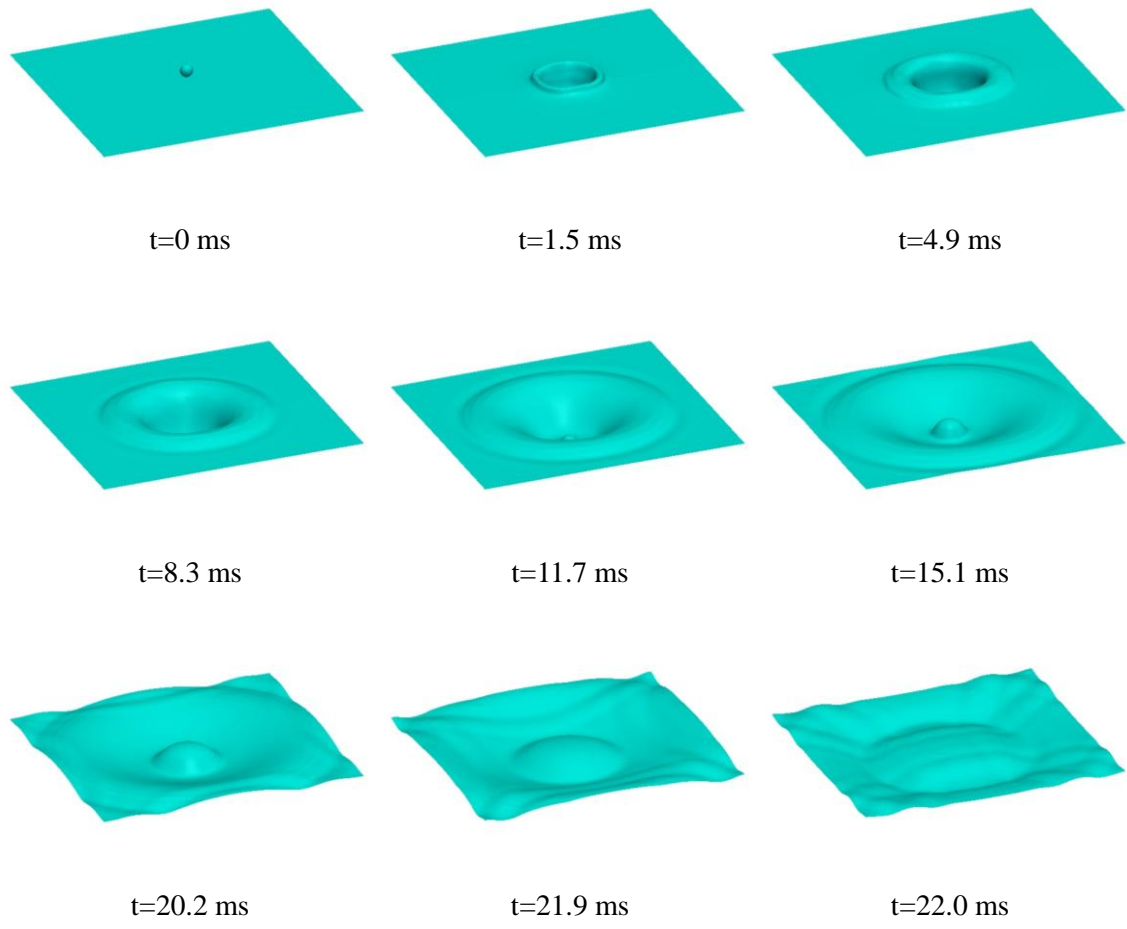

Fig. 7

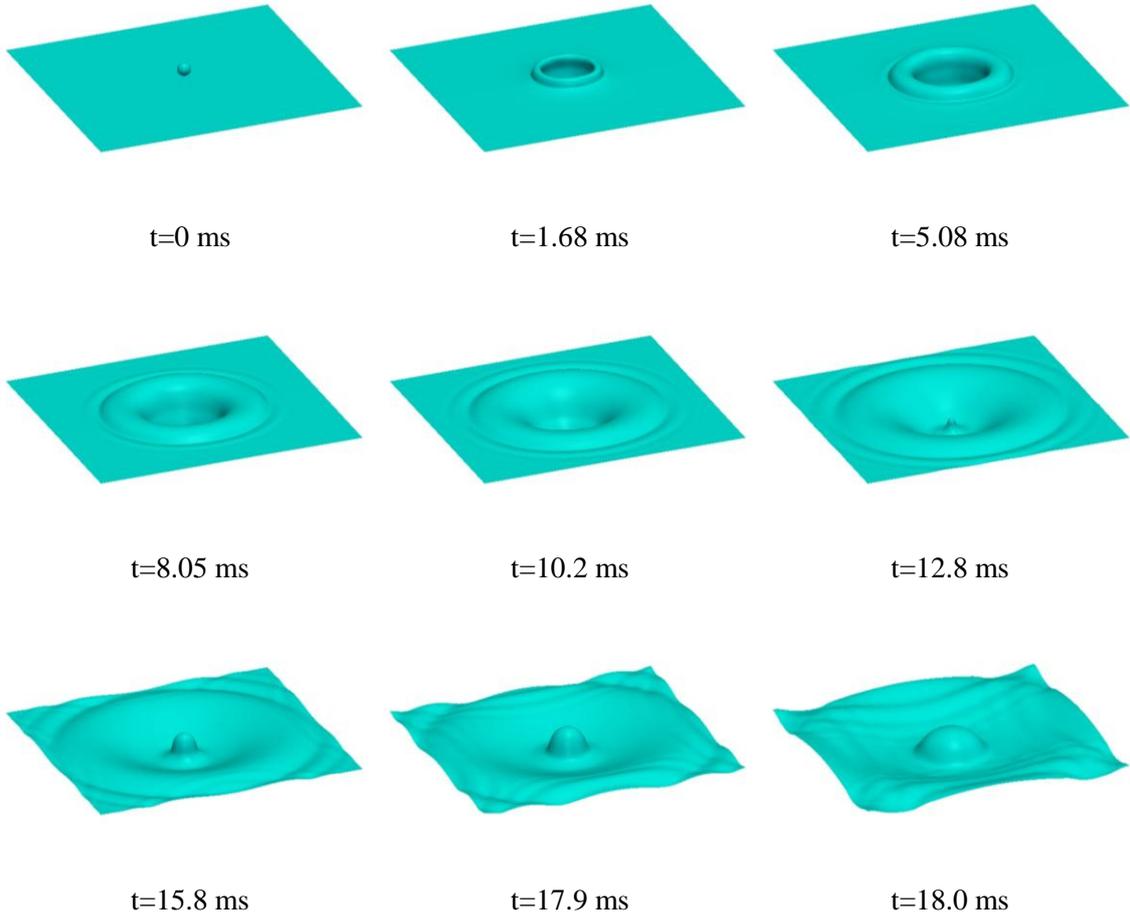

| t=0 ms | t=1.68 ms | t=5.08 ms |
| t=8.05 ms | t=10.2 ms | t=12.8 ms |
| t=15.8 ms | t=17.9 ms | t=18.0 ms |

Fig. 8

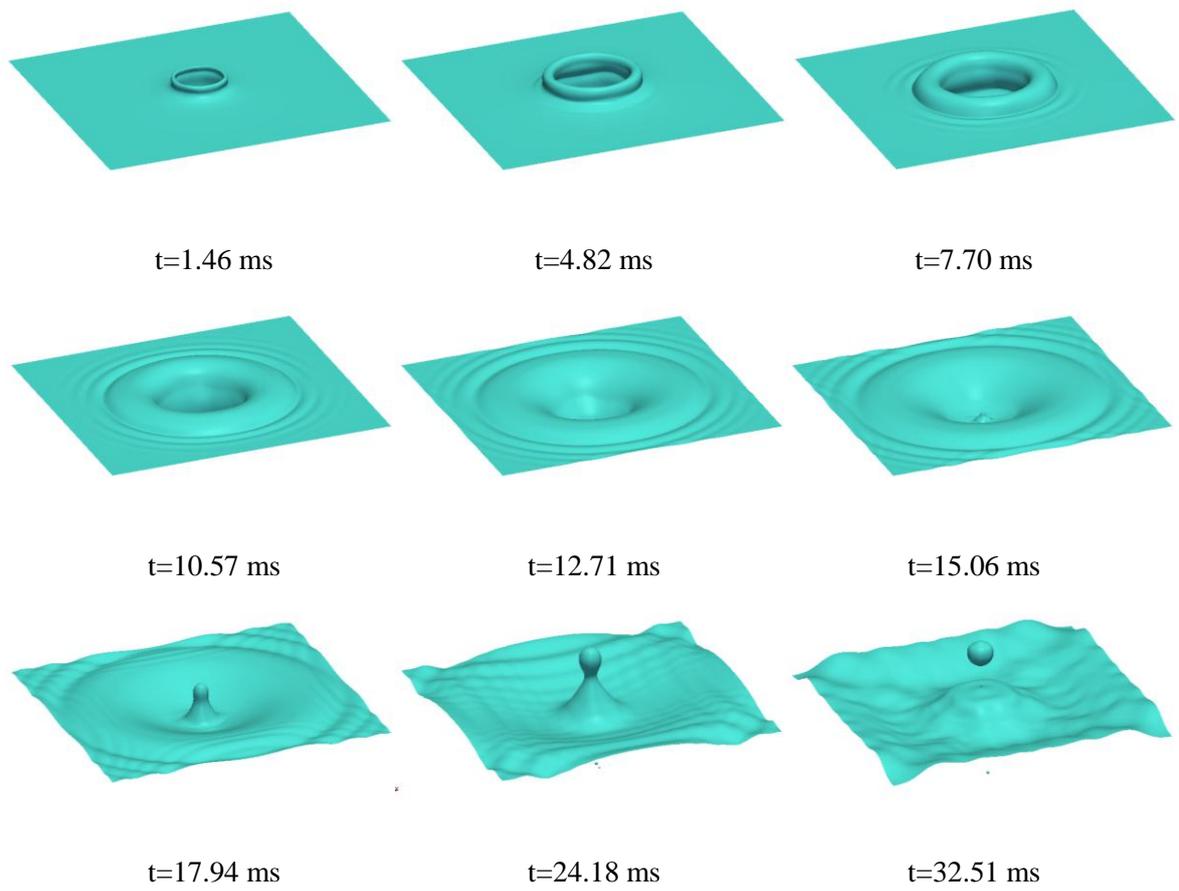

| t=1.46 ms | t=4.82 ms | t=7.70 ms |
| t=10.57 ms | t=12.71 ms | t=15.06 ms |
| t=17.94 ms | t=24.18 ms | t=32.51 ms |

Fig. 9

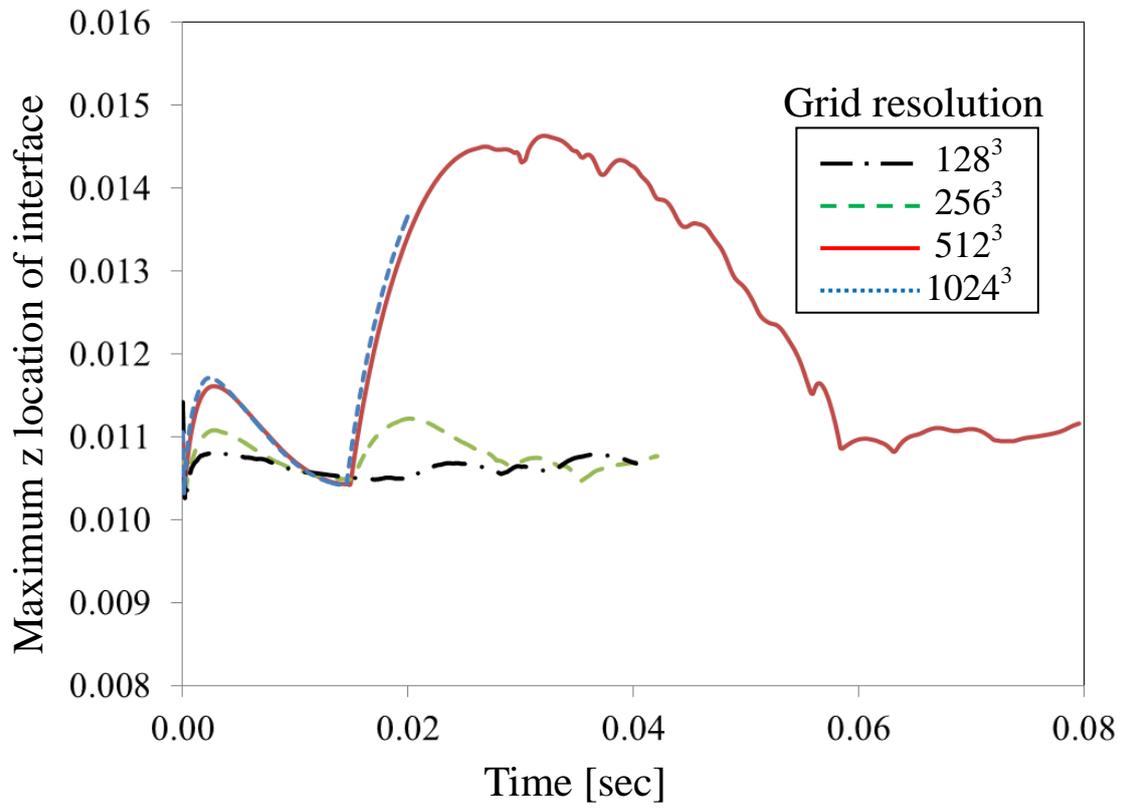

Fig. 10

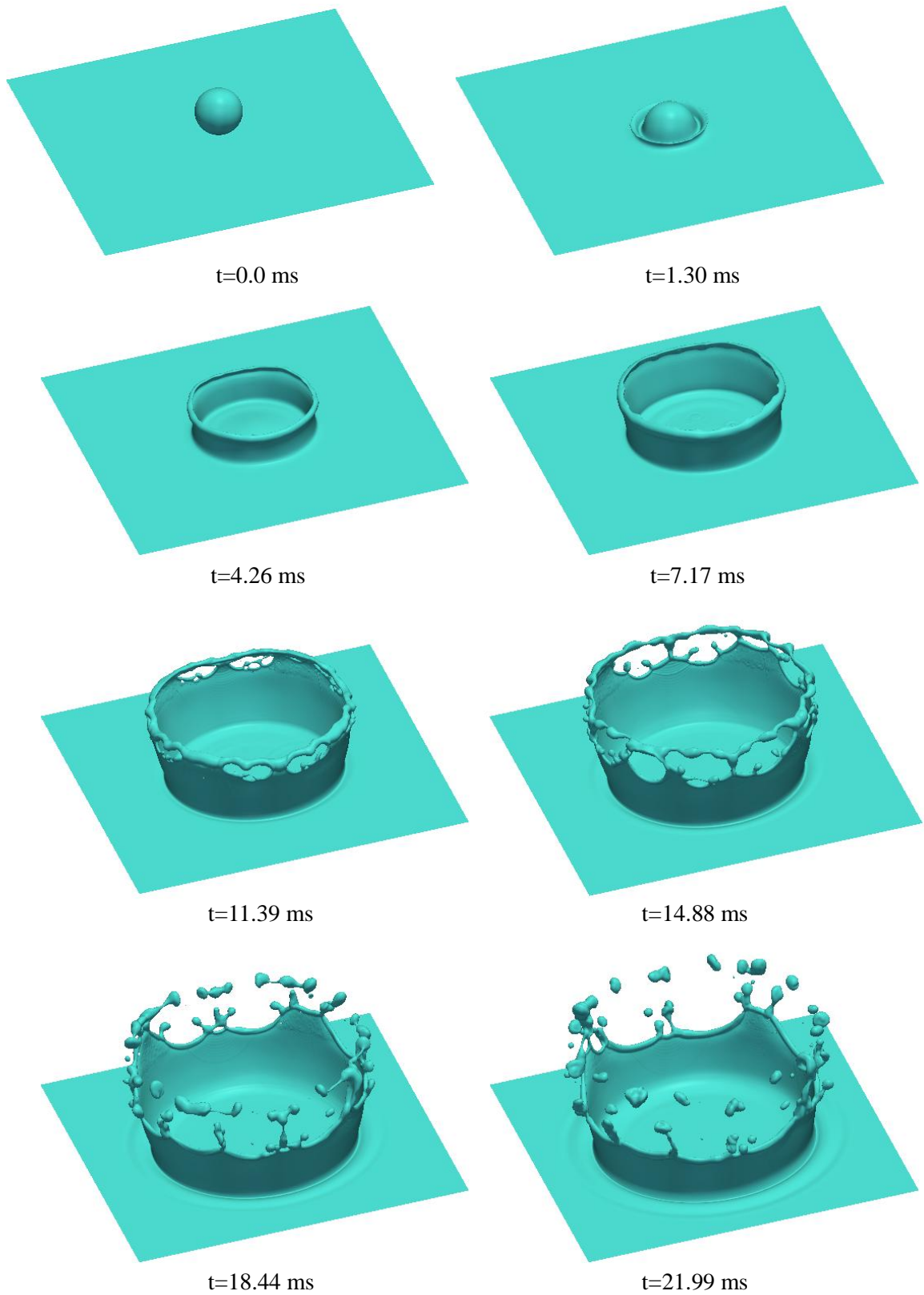

t=0.0 ms  t=1.30 ms

t=4.26 ms  t=7.17 ms

t=11.39 ms  t=14.88 ms

t=18.44 ms  t=21.99 ms

Fig. 11

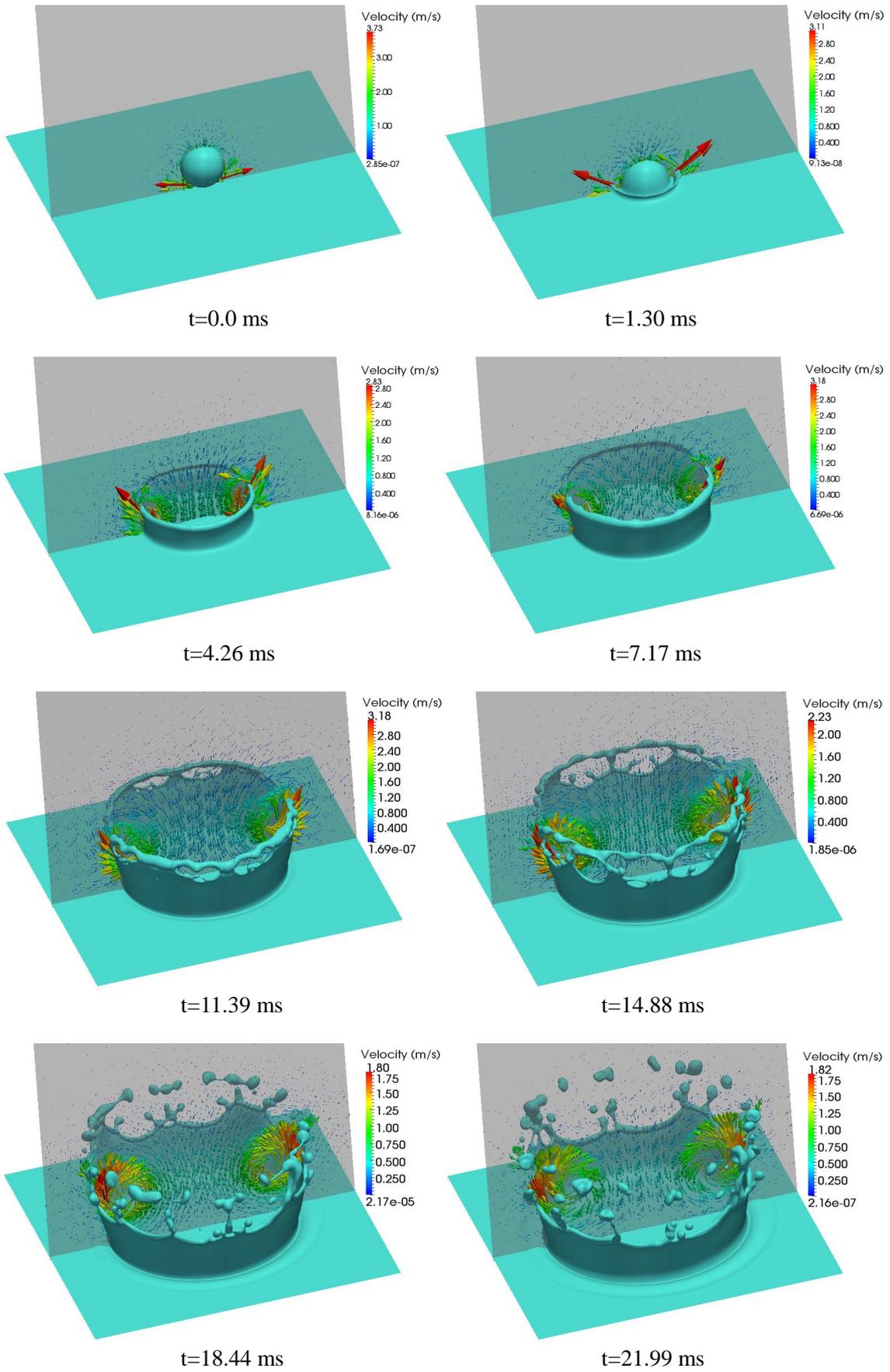

Fig. 12

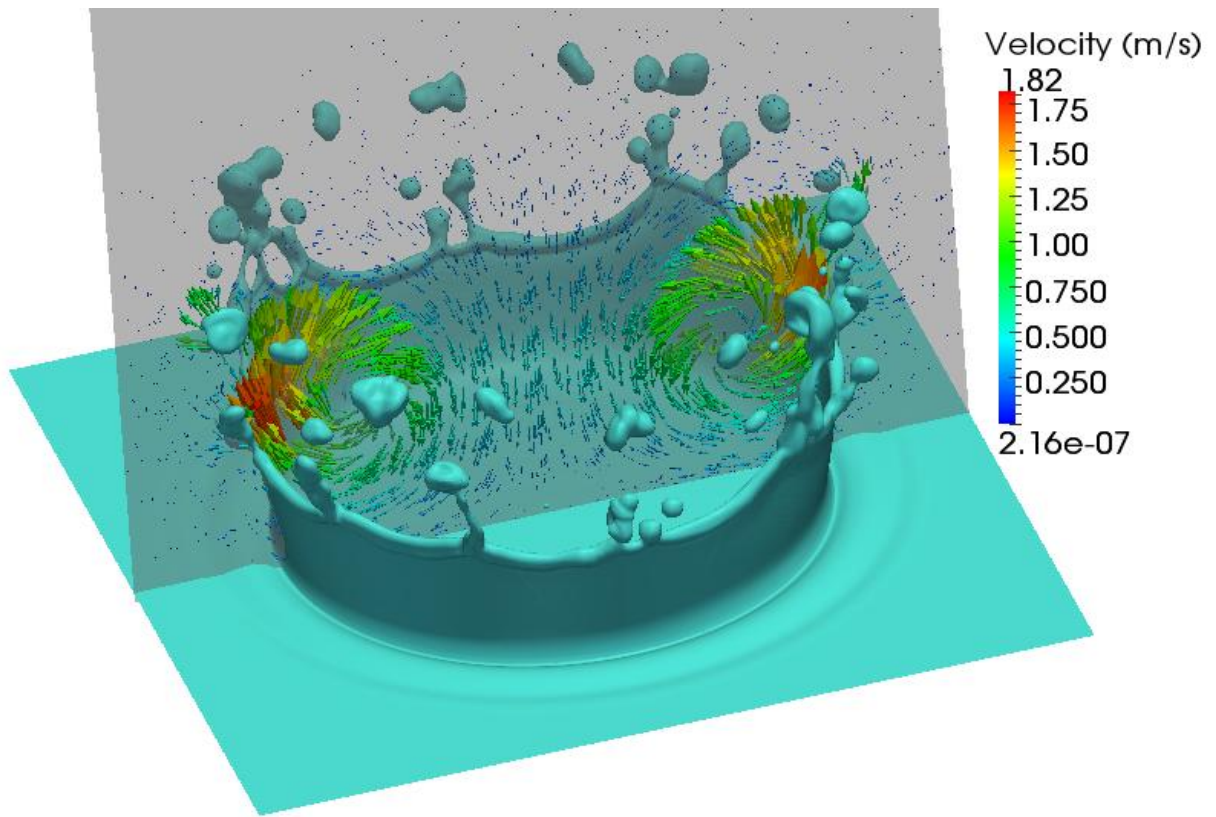

Fig. 13

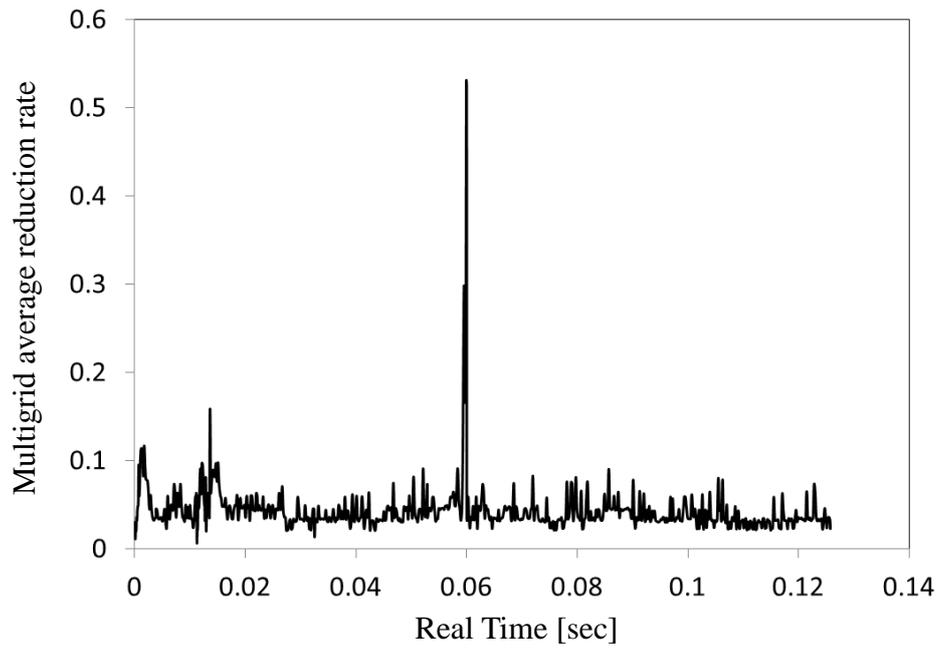

(a)

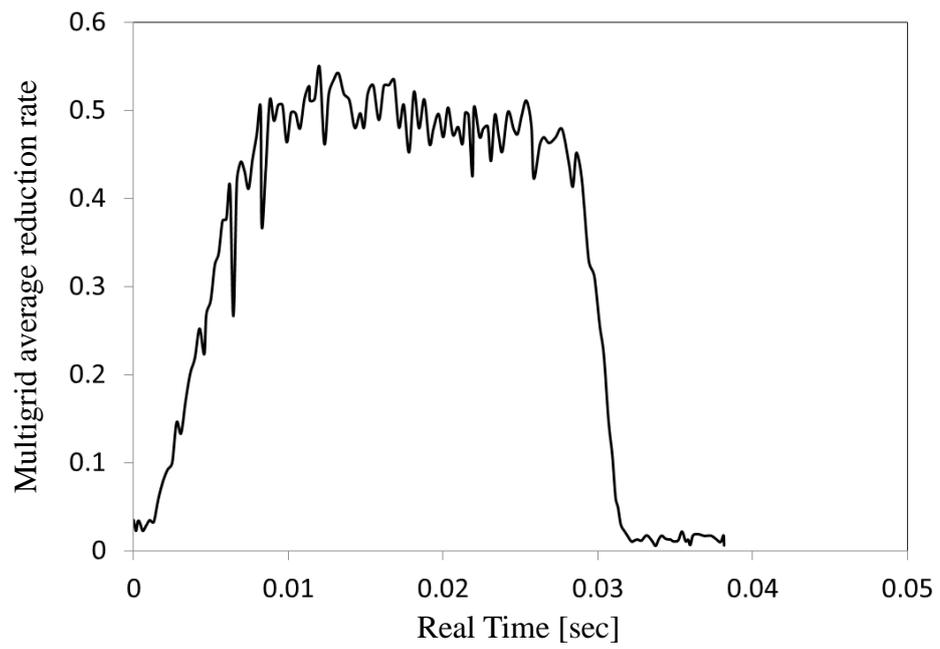

(b)

Fig. 14